# Precision measurements of cross sections of inelastic processes realized in collisions of alkali metal ions with atoms of rare gases

R.A. Lomsadze[1], M.R. Gochitashvili[1], N.O. Mosulishvili[1], and R. Ya. Kezerashvili[2,3]

[1]Tbilisi State University, Department of Exact and Natural Sciences, Tbilisi, 0179, Georgia
[2]New York City College of Technology, The City University of New York, Brooklyn, NY 11201, USA
[3]The Graduate School and University Center, The City University of New York, New York, NY 10016, USA

**Abstract:** A multifaceted experimental study of collisions of $Na^+$ and $K^+$ ions in the energy range 0.5 – 10 keV with He and Ar atoms are presented. Absolute cross sections for charge-exchange, ionization, stripping and excitation were measured using a refined version of the transfer electric field method, angle- and energy-dependent collection of product ions, energy loss, and optical spectroscopy. The experimental data and the schematic correlation diagrams have been employed to analyze and determine the mechanisms for these processes.

*E-mail:* Lomsadze86@hotmail.com

**Introduction**

Existence of the reliable data in slow energy collisions of closed-shell ions with closed shell atoms for various inelastic channels such as the ionization, charge-exchange, stripping and excitation are of considerable interest in atomic physics due to both their importance in fundamental physics and their applications [1]. An accurate determination of the structure of different inelastic cross sections for these collisions is important for understanding the mechanisms for inelastic transitions of the colliding atomic particles. In order to describe quantitatively the excitation mechanism one has to evaluate parameters of the existing theories and explore the contributions of separate inelastic channels for the investigated processes by using experimental measurements.

There are many studies of $K^+$- He, $Na^+$- Ar and $K^+$- Ar collision processes, which have been carried out by various methods [2-9], but available data for the absolute cross section of the above-mentioned processes are contradictory [2,5,6] and in some cases, unreliable [4]. The lack of reliable cross sections data for the processes considered, motivated us to measure the absolute cross sections using a unique combined experimental set up that is incorporated with three collision chambers.

**Experimental technique**

The experimental setup and measurement procedure is discussed in details in Refs. [10, 11-13]. A beam of alkali metal ions ($Na^+$ and $K^+$) from a surface-ionization ion source is accelerated and focused by ion-optics systems. After the beam passes through a magnetic mass spectrometer, it enters the first collision chamber containing He and Ar gasses for measurements of excitation processes in $Na^+$- Ar, $K^+$- Ar and $K^+$- He collisions. The spectral analysis of this radiation was performed in the vacuum ultraviolet (VUV) as well as visible spectral regions. The uncertainties in the absolute excitation cross sections for all mentioned collision system are estimated to be about 20%. Our setup is designed so that when the beam of $Na^+$ and $K^+$ ions passes through the second collision chamber, the measurements of the charge-exchange, ionization and stripping processes occur. These processes are measured by a refined version of the capacitor method [11]. Product particles (secondary positive ions and free electrons) are detected by a collector. A uniform transverse electric field is responsible for the extraction and collection of collision particles. This method [14] yields direct measurements of the absolute ionization and charge-exchange cross sections with the occurrence of 15%. The energy-loss spectrum is obtained by a collision spectroscopy method. The analyzed $Na^+$ and $K^+$ ion beams pass through the third collision chamber by appropriately adjusting the slits prior to entering into a "box" type electrostatic analyzer [12]. The energy resolution of this analyzer is $\Delta E/E = 1/500$. The measured energy-loss spectrum gives detailed information related to the intensity of inelastic processes realized in collision. It is important to emphasize the uniqueness of our experimental approach: the quality of the beam, as well as the experimental conditions for all processes under investigation, always remain identical.

**Experimental results and discussion**

The measured absolute cross sections, in comparison with the results of other authors are presented in Figs. 1-3. In addition, for this processes we have also measured the energies of the electrons emitted in the collisions. It was found that the energy of most liberated electrons is below 17-20 eV. Fig. 4 represents a typical example of the energy-loss spectra for $K^+$- Ar (4a), $Na^+$- Ar (4b) and $K^+$- He (4c). Fig. 5 shows the angular dependence of differential cross sections for $Na^+$- Ar collision. As seen from Figs. 1-3, the characteristic feature for all investigated processes is the magnitude of the cross section. In case when colliding particles have nearly equal masses ($K^+$- Ar cases) the processes are realized effectively, while for asymmetric pairs ($Na^+$- Ar and $K^+$- He) most of these processes have a smaller magnitude of the cross section. Another feature is the difference in the energy dependences. While the ionization cross section (Fig. 2) increases monotonically with the energy, the charge-exchange and excitation cross sections have a complex character (see e.g. curve 4 in Fig. 1 and curve 2 in Fig. 3). Different behavior is observed also for the energy-loss spectrum for $Na^+$- Ar (Fig. 4b). Among the inelastic channels that we could manage to measure is the excitation of Rydberg states of Ar atom with the configuration of $3p^5(^2P)\,4p$ and $3p^5(^2P)\,3d$, though their probability, as seen from Fig. 5, more than one order less compare to the elastic channel. The conclusion that can be drawn from this spectrum is that the energy-loss spectrum has a discrete character. The analysis of the energy-loss spectrum for $K^+$- He collision shows that the excitation of the inelastic channel (one- and two-electron excitation of helium atom and excitation of potassium ion) becomes predominant. The data obtained in this study can be used to draw certain conclusions about possible reasons for these features in the cross section and mechanisms for the corresponding processes. To explain these mechanisms we use a schematic correlation diagram of the adiabatic quasi-molecular terms for this system [15]. This diagrams were presented in our papers [10, 14].

**Charge exchange.** To determine the processes responsible for the charge-exchange in $K^+$- Ar, $Na^+$- Ar and $K^+$- He collisions, we compare the total charge-exchange cross sections presented in Fig. 1 by curve 1, for $K^+$- Ar, by curve 2, for $Na^+$- Ar and by curve 4, for $K^+$- He with those corresponding to the decay of resonance lines presented by curves 5, 3 and 6 respectively. In addition we estimate all the possible reactions those could make a contribution in the total charge exchange processes. We thus can conclude that the charge-exchange processes are governed primarily by the capture of an electron in the ground state of an incident ion, accompanied by the formation of target ion also in the ground state. The analysis of the correlation diagrams shows that a process can occur, as a result of the direct pseudo crossing of the term corresponding to the initial state with the ground state of the system, and $\Sigma$- $\Sigma$ transitions play a dominant role in the charge-exchange processes.

**The ionization.** It follows from the results of the present measurements of emitted electrons in $K^+$- Ar, $K^+$- He, $Na^+$- Ar collisions that the liberation of slow electrons (with the energies less than 17-20 eV) is a characteristic of ionization. In order to determine the channel and mechanism of ionization, we estimate the contribution of several inelastic processes that result in the emission of slow electrons. The contribution of the direct ionization is estimated following Ref. [16]. Estimates of the cross section show that in case of $K^+$- Ar collision this contribution is less than 1%, for $K^+$- He less than 5.5% and for $Na^+$- Ar less than 10%. The double ionization of He and Ar atoms and the capture accompanied by the ionization of target atoms evidently makes a small contribution in $Na^+$, $K^+$ - He, Ar collision because of the large energy defect for these processes and the absence of pseudocrossings of the corresponding quasimolecular terms with the ground state term [10,14]. Consequently, we draw the conclusion that the main mechanisms for $K^+$ - Ar collision pair associated with $\Sigma$-$\Sigma$ transition. The mechanism of ionization for $Na^+$ - Ar collision connected with the promotion of the diabatic term in the continuum. In case of $K^+$ - He collisions we find that the ionization may be caused primarily by the decay of quasimolecular autoionization states. As it seen from energy-loss spectrum in Fig. 5d these could be states with two excited electrons of a He atom ($2s^2$; $2s2p$; $2p^2$) which undergo quasimolecular decay with a high probability. This conclusion follows also from the analysis of the correlation diagram of the system shown in Ref. [16]. Since the ground state term has the symmetry $^1\Sigma$, the states with two excited electrons are populated as a result of $\Sigma$-$\Pi$ and $\Sigma$-$\Pi$-$\Delta$ transitions, associated with the rotation of the internuclear axis along the different nuclear trajectories.

**The stripping.** To determine the mechanism responsible for the stripping processes during the $K^+$- He collisions (stripping in case of $K^+$- Ar and $Na^+$- Ar collisions were not observed) we analyze the correlation diagram [14] and use the procedure described in Ref. [16]. According to our estimation, the governing mechanism for the stripping processes is the transition of the adiabatic term into the continuum.

**The excitation.** For the interpretation of the mechanism responsible for excitation processes in $K^+$- He collision (curve 6 in Fig. 1 and curve 3 in Fig. 3) it is expected to use the correlation diagram of molecular states of the $(KHe)^+$ system [14] and energy-loss spectrum shown in Fig 5d. The term corresponding to the K(4p) does not have immediate crossing point with the entrance term. However, it may be populated from that corresponding to K(3d) excited states by means of a double rotational transition $\Sigma$-$\Pi$-$\Delta$ between the entrance term and that corresponding to K(3d) state via the cascade transition, from the level 3d to 4p. To analyze the results for the excitation of argon atom into 4s and 4s$^/$ states and for the capture of an electron into the resonance 3p state of sodium atom for the $Na^+$- Ar collision we will use the schematic correlation diagram. The analysis shows that excitation of the argon atom lines ($\lambda$ =104.8 nm and $\lambda$= 106.7 nm, transition 4s$^/$, 4s - 3p) revealed to the oscillatory structure (curve 1 in Fig. 3). The observed oscillatory structure, a comparatively small cross section ($\sigma = 10^{-18}$ cm$^2$) and large oscillating depth (see curve 1 in Fig. 3) indicate, according to [17], that the contribution of the rotational $\sum$-$\prod$ transition with population of the 4p energy level of an Ar atom is significant for the excitation processes. These oscillations are due to the interference of energetically closed quasi-molecular states of the Na$^+$-Ar (4p) and Na (3p)-Ar systems. However, in accordance with the interference model [18] the energy dependence of the excitation cross section for Na (3p) (curve 3 in Fig.1) may exhibit an oscillatory structure in antiphase with the dependence observed for the lines of the Ar atom. The energy dependence of the excitation cross section of a Na atom ($\lambda$ = 589. 0 – 589.6 nm, 3p-3s transition) exhibits only a small structural singularity. It was found that the absence of a clearly manifested oscillatory structure on the excitation cross section of the lines of the Na atom can be associated with the effect of cascade transition from the $4s^2\ ^2S_{1/2}$ and $3d\ ^2D_{3/2}$ to the 3p level.

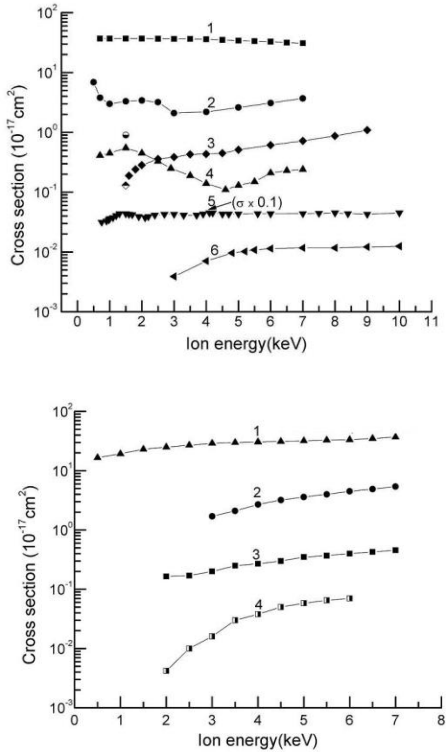

FIG.1. The charge exchange cross-sections of K$^+$ and Na$^+$ ions collision with He and Ar atoms. Curves: (1), (2) and (4) – charge exchange in ground (4s), (3s) and (4s) states for K$^+$- Ar, Na$^+$- Ar and K$^+$- He collision respectively. Curves: (3), (5) and (6) – charge exchange in resonance (3p), (4p) and (4p) states for Na$^+$- Ar, K$^+$- Ar and K$^+$- He collision respectively. ◐,◊ - charge exchange and charge exchange with excitation of target ion for Na$^+$- Ar from Ref. [9].

FIG.2. The absolute ionization cross-sections. Curves: (1) for K$^+$- Ar collision; (2) for Na$^+$- Ar collision; (3) for K$^+$- He collision; (4) the formation of K$^{2+}$ product ions (stripping) in K$^+$- He collision.

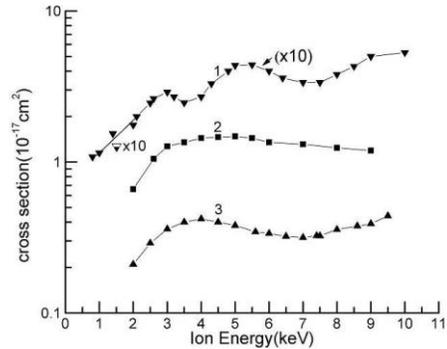

FIG.3. The excitation function for Ar and He target atoms and potassium ion in Na$^+$- Ar and K$^+$- He collisions. Curves: (1) excitation of Ar atom in (4s$'$;4s) states for Na$^+$- Ar collision; ▼- excitation of Ar atom for Na+ -Ar [9]; (2) excitation of K$^+$ ion into the 4s$'$ state for K$^+$- He collision; (3) excitation of He atom in (2p) state for K$^+$- He collision.

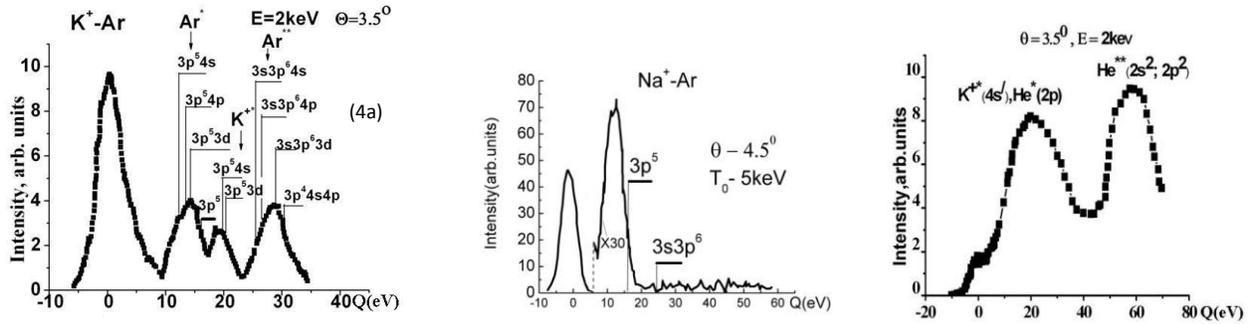

FIG.4. Energy-loss spectra. (a) The energy-loss spectrum for $K^+$- Ar collision. (b) The energy-loss spectrum for $Na^+$-Ar collision. (c) The energy-loss spectrum for $K^+$- He collision.

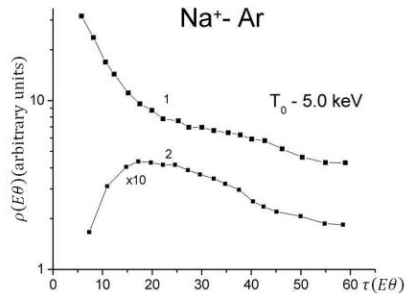

FIG.5. Differential cross-section in collision of $Na^+$ ion with Ar atom. Curves: (1) elastic scattering; (2) excitation of Rydberg states of Ar with the electron configuration $3p^5(^2P)4p$ and $3p^5(^2P)3d$.

## Acknowledgments

This work was supported by the Georgian National Science Foundation under the Grant No.31/29 (Reference No. Fr/219/6-195/12). R.Ya.K. and R.A.L gratefully acknowledge support from the International Research Travel Award Program of the American Physical Society, USA.